The Thermophysical Properties of the Bagnold Dunes, Mars: Ground-truthing Orbital Data

Christopher S. Edwards[1], Sylvain Piqueux[2], Victoria E. Hamilton[3], Robin L. Fergason[4], Ken E. Herkenhoff[4], Ashwin R. Vasavada[2], Kristen A. Bennett[1], Leah Sacks[5], Kevin Lewis[6], Michael D. Smith[7]

[1]Northern Arizona University, Department of Physics and Astronomy, NAU BOX 6010 Flagstaff, AZ 86011, USA, Christopher.Edwards@nau.edu

[2]Jet Propulsion Laboratory, California Institute of Technology, Pasadena, CA, USA

[3]Southwest Research Institute, Boulder, CO, USA

[4]U.S. Geological Survey, Astrogeology Science Center, Flagstaff, AZ

[5]Carelton College, Northfield, MN, USA

[6]Johns Hopkins University, Baltimore, MD, USA

[7]NASA Goddard Space Flight Center, Greenbelt, MD, USA

Key Points:

1) Thermally derived particle sizes of the Bagnold dunes from orbit and landed assets are consistent with direct grain size measurements

2) Thermally derived particle sizes are not dramatically affected by surface ripples or thin layers of induration/armoring

3) Sub-pixel mixing of sand with nearby materials likely resulted in overestimated particle sizes in previous orbital measurements

Abstract: 244 words

Body Text + Figures: 8026

Body References: 78

Figures: 11

Tables: 1

*1 Abstract*

In this work, we compare the thermophysical properties and particle sizes derived from the Mars Science Laboratory (MSL) rover's Ground Temperature Sensor (GTS) of the Bagnold dunes, specifically Namib dune, to those derived orbitally from Thermal Emission Imaging System (THEMIS), ultimately linking these measurements to ground-truth particle sizes determined from Mars Hand Lens Imager (MAHLI) images. In general, we find that all three datasets report consistent particle sizes for the Bagnold dunes (~110-350 μm, and are within measurement and model uncertainties), indicating that particle sizes of homogeneous materials determined from orbit are reliable. Furthermore, we examine the effects of two physical characteristics that could influence the modeled thermal inertia and particle sizes, including: 1) fine-scale (cm-m scale) ripples, and 2) thin layering of indurated/armored materials. To first order, we find small scale ripples and thin (approximately centimeter scale) layers do not significantly affect the determination of bulk thermal inertia from orbital thermal data determined from a single nighttime temperature. Modeling of a layer of coarse or indurated material reveals that a thin layer (< ~5 mm; similar to what was observed by the Curiosity rover) would not significantly change the observed thermal properties of the surface and would be dominated by the properties of the underlying material. Thermal inertia and grain sizes of relatively homogeneous materials derived from nighttime orbital data should be considered as reliable, as long as there are not significant sub-pixel anisothermality effects (e.g. lateral mixing of multiple thermophysically distinct materials).

## 2 Introduction

The surface of Mars has been characterized using thermal infrared observations from orbit from the time of the Mariner and Viking orbiter missions (e.g. Neugebauer et al., 1971). Viking Infrared Thermal Mapper (IRTM) (Chase et al., 1978) observations (e.g. Christensen, 1982; Christensen, 1983, 1986; Kieffer et al., 1977; Kieffer et al., 1976) permitted the physical characterization of broad regions of the Martian surface to be constrained though little geologic context was available due to the large footprint of the IRTM instrument (~25km). When the Thermal Emission Spectrometer (TES) (Christensen et al., 2001) instrument onboard the Mars Global Surveyor arrived, new views of the surface at much higher resolution (~3x6 km) permitted a host of additional thermophysical investigations, which linked observations to local/regional scale geologic processes (e.g. Bandfield & Edwards, 2008; Bandfield & Feldman, 2008; Mellon et al., 2000; Nowicki & Christensen, 2007; Putzig & Mellon, 2007b; Putzig et al., 2005). However, with the arrival of the Thermal Emission Imaging System (THEMIS) (Christensen et al., 2003) instrument onboard the 2001 Mars Odyssey orbiter and its 100 m/pixel thermal infrared data (Christensen et al., 2004), new investigations were made that linked outcrop-scale compositional information to material physical properties (e.g. Bandfield, 2008; Bandfield et al., 2013; Bandfield & Rogers, 2008; Christensen et al., 2003; Christensen et al., 2005; Edwards et al., 2009; Edwards et al., 2014; Edwards et al., 2008; Edwards & Ehlmann, 2015; Hamilton & Christensen, 2005; Osterloo et al., 2008; Rogers et al., 2009; Rogers & Bandfield, 2009; Rogers et al., 2005; Rogers & Fergason, 2011; Rogers & Nazarian, 2013; Tornabene et al., 2008). These new investigations robustly characterized the geologic histories and origins of surfaces on Mars.

The primary material property under consideration here is thermal inertia (TI), defined as TI = $(k\rho c)^{1/2}$, where k is the thermal conductivity, $\rho$ is the bulk density of the material, and c is specific heat. TI is dominated by the bulk thermal conductivity on Mars (Jakosky, 1986; Kieffer et al., 1973; Presley & Christensen, 1997a, 1997b, 1997c) and can be related to an effective particle size for the upper several milli- to deci-meters of surface material (e.g. Piqueux & Christensen, 2011; Presley & Christensen, 1997b, 1997c; Presley & Craddock, 2006). The conversion from temperature to TI and ultimately to grain size has been applied to both orbital and landed spacecraft data (Fergason et al., 2006a; Kieffer et al., 1973). However, linking these effective derived particle sizes to *in situ* measurements (i.e., ground-truthing) by correlating derived thermophysical properties of surface materials with particle size distributions derived from imaging presents a significant challenge, due to disparate instrument resolutions as well as sampling cadence and times. This is especially true for those surfaces for which TI values which are not consistent with loose particulate material (<~350 J $m^{-2}$ $K^{-1}$ $s^{-1/2}$; e.g. Piqueux & Christensen, 2009a, 2009b) or igneous bedrock (>1200 J $m^{-2}$ $K^{-1}$ $s^{-1/2}$; e.g. Edwards et al., 2009). These intermediate thermal inertia regimes (e.g. ~350-1200 J $m^{-2}$ $K^{-1}$ $s^{-1/2}$) can be further complicated by the non-homogeneity of the surfaces in question, which can introduce significant sub-pixel anisothermality into a single temperature measurement (e.g. Putzig & Mellon, 2007a, 2007b). Additionally, the limited temporal sampling of most fixed-local-time polar orbiting missions (e.g., those carrying TES and THEMIS) complicates the interpretation of thermophysical signatures. The temporally limited measurements obtained by these instruments do not capture the shape of the diurnal temperature response curve (e.g.

Fergason et al., 2006b), though a single nighttime temperature measurement can still provide a strong lever arm to disambiguate the physical properties of the surface. Importantly, over the lifetime of the Mars Odyssey mission the orbit migrated in local time sampling from ~0300-0700 and 1500-1900, which provided additional diurnal coverage over much of the martian surface.

The Mars Science Laboratory (MSL) Curiosity rover spent approximately 20 sols (1222-1242) exploring and characterizing an active dune field in Gale Crater informally named the Bagnold Dunes. Specifically, the rover was parked at a barchan dune informally named "Namib" over these sols. The Bagnold dune field is located on the northwest flank of Aeolis Mons (Fig. 1a), informally known as "Mt. Sharp", and provides an excellent opportunity to link orbitally derived thermophysical properties to *in situ* observations, for several reasons:

1) the active nature of this dune field limits the likelihood of cementation of (and therefore vertical variation in) the particulate material;

2) aeolian dunes likely represent a relatively homogeneous material, reducing the potential for material mixture induced anisothermality of the surface within a given footprint;

3) the large areal extent of the dune field which is resolved by many THEMIS pixels (100m/px); and

4) these dunes have particle sizes large enough (>35μm) to be characterized by imaging instruments (e.g. ChemCam Remote Microscopic Imager (ChemCam RMI) (Maurice et al., 2012; Wiens et al., 2012), and Mars Hand Lens Imager (MAHLI) (Edgett et al., 2012)) onboard Curiosity.

In this work we link the orbital THEMIS thermophysical characteristics (and their interpretation in terms of particle sizes) with ground observations using the thermal infrared measurements of the Rover Environmental Monitoring Station (REMS) (Gómez-Elvira et al., 2012) – Ground Temperature Sensor (GTS) (Hamilton et al., 2014; Sebastian et al., 2010) to robustly determined particle sizes from Curiosity rover imagery. This inter-dataset linkage can be used to better understand dunes and other landforms across the surface of Mars and aid in interpretations of aeolian and geological processes.

3  *Methods*

3.1 THEMIS Observations

THEMIS has been in operation around Mars onboard the 2001 Mars Odyssey spacecraft for ~15 years. Over this timeframe, it has imaged the majority of the planet multiple times, both day and night (Edwards et al., 2011b), and has acquired many observations of Gale Crater (Hamilton et al., 2014). THEMIS data were processed through standard techniques that include the removal of time-dependent focal plane temperature drift (Bandfield et al., 2004), temperature variations across the THEMIS calibration flag (Edwards et al., 2011b), map projection and then the removal of band-dependent and band-independent row- and column-correlated noise (Edwards et al., 2011b; Nowicki et al., 2013). These processing techniques result in noise-corrected, calibrated radiance data. THEMIS brightness temperatures were determined by matching the expected radiance from a uniformly emitting (blackbody) source to the THEMIS-measured radiance at band 9 (Bandfield et al., 2004; Christensen et al., 2004; Edwards et al., 2011b). The brightness temperature at band 9 (12.57µm) is taken to represent the surface temperature (Fergason et al., 2006b) because the Martian atmosphere is relatively transparent in that band and the highest signal to noise measurements are obtained with this band, especially for nighttime data (Christensen et al., 2004; Edwards et al., 2009; Edwards et al., 2011b; Fergason et al., 2006b). When deriving thermophysical properties, nighttime data (typically ~3-6 AM for the Mars Odyssey orbit) are used as they minimize the effects of slopes and albedo, permitting a more reliable TI determination from a single data point. In deriving THEMIS thermal inertia values for this work (Figure 1B), the KRC thermal model (Kieffer, 2013) was employed. We used this model with a host of input parameters such as the albedo, atmospheric opacity, elevation (and corresponding surface pressure scaled based on the seasonal evolution observed by the Viking Landers), emissivity, surface geometry (i.e., slope and azimuth), temperature dependent material properties (specific heat and thermal conductivity) and modeled temperatures for a specified seasonal and local time range and interval for the region of interest on the surface (see Kieffer, 2013 for a comprehensive overview of the model capabilities). Previous work using the KRC thermal model with THEMIS data determined these input parameters on a framelet (256-line segment parts of the THEMIS image) basis (Fergason et al., 2006b) to facilitate the production of THEMIS thermal inertia as a global dataset. For this study we have developed a method to derive the thermal inertia using higher resolution input parameter datasets available at the MSL landing site on a pixel by pixel basis (Figure 2B). This is similar to several other investigations (e.g. Catling et al., 2006; Sefton-Nash et al., 2012), although those studies did not use the publicly available KRC thermal model. Furthermore, we use the "full" KRC model to develop a temperature to thermal inertia lookup table with 40 thermal inertia values ranging from 10-2200 J $m^{-2}$ $K^{-1}$ $s^{-1/2}$, which runs for two martian years prior to outputting data in the third year to achieve seasonal stability. This is in contrast to previous work that relied on the one-point mode which only runs a subset of seasonal cases (Fergason et al., 2006b; Kieffer, 2013). Critically, our updated method, while computationally inefficient, takes advantage of spatially registered data at the same scale as the THEMIS IR data. Data that was input into our model included a Context Camera (CTX) (Malin et al., 2007) Digital Terrain Model (DTM) used for the slope, azimuth, and elevation (Figure 3B-D) of the individual pixel as well as THEMIS visible Lambert albedo (V01494002; Edwards et al., 2011a) that is tied to TES Lambert albedo (Figure 3A). We assumed a surface emissivity of 0.98 (relevant for Gale Crater (Bandfield et al., 2000)) and we included latitude, longitude, and local time backplanes. The method described above was applied to THEMIS image I17950012. This image was converted to band 9 brightness temperature and resulted in 10s of millions of KRC model runs that generated temperature to thermal inertia lookup tables with 40 thermal inertia nodes on a pixel by pixel basis (~300K pixels). The look up table was then applied using linear interpolation to the THEMIS temperatures. THEMIS image I17950012 has a local time of 04:12, a solar longitude of 349.1°, and a visible atmospheric dust opacity of 0.4 that was scaled to 6 mbar (Smith et al., 2016) and was acquired in Mars year 27. This image was chosen due to its ideal coverage of the site under investigation (e.g. the entire rover path present and future) as well as its overall calibration quality (no data dropouts, limited time between the shutter closing image and data acquisition, lack of saturated or undersaturated pixels, and no evidence of enhanced line-line noise) (Bandfield et al., 2004; Christensen et al., 2004; Edwards et al., 2011b).

Following the conversion to thermal inertia, THEMIS data were transformed to particle size following the relationships established by (Presley & Christensen, 1997b, 1997c) and refined by (Piqueux & Christensen, 2011). This method uses a parameterization of the model results of Piqueux and Christensen (2011) and accounts for an input surface pressure (scaled for the elevation and season, ~800 Pa for Gale Crater), average upper layer skin depth temperature (180K), thermal conductivity derived from thermal inertia, porosity reasonable for particulate materials (40%; Denekamp & Tsur-Lavie, 1981), a solid density of 2950 kg/m$^3$ (consistent with basalt), and specific heat for basalt (595.5 J kg$^{-1}$ K$^{-1}$; Fujii & Osako, 1973). We limit the scale of particle sizes to 1mm in Figure 2C because surfaces with effective particle sizes greater than this can no longer be interpreted as strictly particulate media (Piqueux & Christensen, 2009a, 2009b) and interpretations become non-unique. Effective particle sizes larger than 1 mm are consistent with cemented or indurated materials and without further context are non-unique. While further information can certainly be drawn from the larger effective particle sizes, the focus of this paper is the Bagnold sand dunes, and as such the need to consider coarser particle sizes was not warranted.

3.2 GTS Observations

The GTS observations of interest span Sols 1222 through 1242 and were taken while the Curiosity rover was parked with the GTS sensor pointed at the west-facing, secondary slip face of Namib dune (Figure 4A). GTS data with the best calibration were selected and grouped by local time bins (Figure 4B) to match the output of the KRC thermal model used for analysis. Several measurement uncertainties exist when considering the GTS data, including potential shadowing by the MSL mast within the GTS observation footprint, and the influence of the radioisotope thermoelectric generator (RTG). The effects of these uncertainties are likely difficult to quantify but may contribute to up to a ~4K temperature error (Hamilton et al., 2014; Martínez et al., 2014). A series of diurnal temperature curves were generated for a range of thermal inertia values (varying from 100-350 J m$^{-2}$ K$^{-1}$ s$^{-1/2}$ by 10 J m$^{-2}$ K$^{-1}$ s$^{-1/2}$ increments), albedo values (from 0.07-0.2 by 0.01 increments) with all the other appropriate inputs (e.g. REMS-derived UV opacities scaled to 6.1 mbar resulting in 0.337 (Smith et al., 2016), latitude, longitude, elevation, local time, assumed surface emissivity of 0.98, temperature dependent material properties, etc.) using the KRC thermal model, for comparison with measured temperatures from GTS data. Several metrics to assess the quality of the fit of the GTS temperature data to the KRC model-derived output over the 21 sols were computed including: 1) the RMS difference between the GTS and modeled data, 2) the difference between the minimum and maximum diurnal temperature (ΔT) of GTS and modeled data, and 3) the difference of average diurnal temperatures between the GTS and modeled data. We take three approaches to determine the albedo and thermal inertia in order to compare methods. The first is a straightforward best fit by "eye", where the best fit is qualitatively estimated. Second, the minimum RMS difference is used to simultaneously solve for albedo and thermal inertia. And third, we use the minimum difference of average diurnal temperatures to solve for albedo and then, using the derived albedo, we find thermal inertia by using the minimum ΔT. The functional effect of the thermal inertia is to change the amplitude of the diurnal curve, while albedo, which controls the total amount of energy absorbed by the surface, results in a shift of the average temperature of the diurnal curve (Figure 6). Each of these thermal inertia values was then converted to particle size following the methods described above for THEMIS data. This thermally derived albedo (0.08-0.13, Figure 5) is compared to orbitally derived

Lambert albedo from THEMIS visible imagery (0.15 is typical for the Bagnold dunes) to ensure that a reasonable albedo is derived.

Several modeling approaches were taken in order to assess the sensitivity of the derived thermal inertia to armoring/layering and variations in slope and azimuth in the GTS footprint (Figure 5B & C). There are two slope peaks in the GTS footprint at ~7.5° and ~28° and azimuths range from 225-300° (Figure 5D), which potentially indicate that a simple planar fit is not sufficient. The three approaches taken are as follows:

1) A best fit plane over the GTS footprint (Figures 4 & 5) is calculated and used as the slope and azimuth input to the thermal model.

2) Each scene within the GTS footprint was discretized into ~5,000 individual slope and azimuth combinations (facets) derived from Navcam stereo data. The resulting diurnal temperatures were modeled for each thermal inertia and albedo combination, using the additional inputs described above. The sampling of these facets was calculated by evenly filling the GTS FOV in angular space rather than in rectified spatial coordinates and intersecting these vectors with the Navcam surface mesh, selecting the slope and azimuth combination. This provides a reasonable approximation for the weighting of the near- and far-field contributions to the measured radiance. After modeling ~5,000 unique diurnal curves, representing the range of temperatures experienced by the surface for the given slope and azimuth distributions (Figure 5), these temperatures are converted into radiance, averaged, and converted back to a single brightness temperature per local time bin, and allowed us to capture the effect of slopes at sub-pixel scales. This process results in set of diurnal curves as a function of thermal inertia and albedo that can then compared to the GTS data. While the resolution of the thermal inertia and albedo lookup table used is somewhat coarse – increments of 10 J m$^{-2}$ K$^{-1}$ s$^{-1/2}$ and 0.01 units of Lambert albedo – these incremental values result in temperature variations that are likely within the uncertainty of the GTS instrument (less than a few K).

3) Other scenarios were considered that match the fine-scale imagery of the dune scuffs by the rover wheel. These images indicate a slightly indurated, thin upper layer that has a lag of coarser particles. We model this scenario using a layered thermal model in which the upper material thermal inertia, lower material thermal inertia, and upper layer thickness can be varied independently. Due to the non-unique results of this method, the determination of the best fit was achieved by eye. We used a set of input variables (e.g. layer thicknesses etc.) that included those derived from the simple and facet models described above. The layered model is used in conjunction with image derived particle sizes converted to thermal inertia (185 J m$^{-2}$ K$^{-1}$ s$^{-1/2}$ and 150 μm for the lower material and 225 J m$^{-2}$ K$^{-1}$ s$^{-1/2}$ and 350 μm for the upper material) to simply assess the effects small scale armoring/induration of the upper most layers has on the thermophysical properties of dunes.

3.3 Particle sizes from Imagery

MAHLI data spanning the campaign of the Bagnold dunes (Table 1) were used to determine the true particle sizes of the dune in question. Because each whole GTS footprint was not analyzed with fine-scale rover imagery, given the relatively large GTS footprint (meters scale), a set of representative MAHLI images were chosen for analysis. These images cover the lee and stoss sides of a small-scale ripple. They also capture both disturbed and undisturbed materials that may reveal armoring of finer grained materials by a coarser grained surficial component (observed at MSL's "Rocknest" site and by other missions; e.g. MER (Fergason et al., 2006a)), or surface crusts formed by weakly cemented particulate materials (e.g. Piqueux & Christensen, 2009a, 2009b). Two methods were employed to derive the particle size from MAHLI data (Figures 7 & 8) which encompassed a range of standoff distances and thus resolutions:

1) Digital particle size analysis using a power spectral density function of particle sizes determined using Morlet wavelets (Buscombe, 2013). This method requires fully resolved particles and appropriate choice of input parameters including sampling density, the relationship between the image size (e.g. number of pixels) and the relative size of the particles (in pixels), which ensures that there are sufficient points to analyze and that there are sufficient octaves and notes to capture the size distribution of the particles, respectively (Buscombe, 2013). The result of this algorithm is a particle size distribution and the average particle size ± the standard deviation (Table 1; Figure 8).

2) For a subset of MAHLI images (Figure 7), point counts were conducted by hand. In this methodology, a grid of points was spaced evenly that resulted in at least 100 measurement points was overlain on the image. The long and short axes of the particle that lands under the given crosshair of a given grid were measured and their average is reported as the particle's size. Crosshairs that did not align with a particle were excluded from the point count survey and crosshairs that align with an unresolvable particle are lumped into a category that is defined by the minimum resolvable particle of the image (3x the image resolution). The average and standard deviation of these particle sizes are then determined from this distribution (Table 1; Figure 7).

*4   Results*

4.1 Particle sizes from Orbit

THEMIS data provide valuable medium resolution information about the effective particle sizes of surfaces across vast regions of Mars at scales relevant for local studies. These data leverage the thermophysical history and techniques of the Viking IRTM (e.g. Christensen, 1983; Edgett & Christensen, 1991; Kieffer et al., 1977) and TES (e.g. Jakosky et al., 2000; Mellon et al., 2000; Putzig & Mellon, 2007a; Putzig et al., 2005) instruments. The particle size of aeolian dunes are of particular interest for several reasons for this work. First, aeolian dunes provide the most uniform (from a particle size perspective) surfaces to compare to landed observations in both visible imagery and thermally (e.g. Fenton & Mellon, 2006; Fergason et al., 2006a; Fergason et al., 2006b) and have a predictable grain size constrained by wind tunnel experiments under Martian pressures (e.g. Greeley et al., 1980). Second, the Bagnold dunes represent the first time that an active dune, rather than a mantled or inactive bedform, can be directly assessed from orbit and ground. The Namib dune, in the southwest region of the main Bagnold dune field where Curiosity stopped does not fill an entire THEMIS pixel, which would also contain radiance from rock of the Murray and Stimson formations. This results in mixing and anisothermality within a THEMIS pixel, ultimately reducing the utility of this area for direct comparison with THEMIS. However, the remainder of the Bagnold dune field, which fills many THEMIS pixels, is available for use and has similar characteristics to the Namib dune. There is no indication that Namib dune is meaningfully different than the remainder of the field from a particle size perspective and likely has a similar sediment source and transport history. It is worth mentioning that slightly stronger olivine-spectral features are observed in association with the barchan dunes, potentially suggesting minor differences in sorting may be present (Laporte et al., submitted; Viviano-Beck et al., 2014), though these minor spectral differences are difficult to quantity from a physical properties perspective.

From analysis of THEMIS data (I17950012), the main Bagnold dune field (observed by multiple THEMIS pixels) has a minimum apparent thermal inertia of 200 J $m^{-2} K^{-1} s^{-1/2}$, a maximum of 310 J $m^{-2} K^{-1} s^{-1/2}$ and an average thermal inertia of 240 ± 20 J $m^{-2} K^{-1} s^{-1/2}$. This corresponds to a minimum particle size of 148 μm, a maximum of 968 μm and an average particle size of 251±84 μm. A single porosity for the entire image was used (40%) and is likely relevant for the regolith dominated surfaces (e.g. Denekamp & Tsur-Lavie, 1981; Piqueux & Christensen, 2011). The average particle size values were calculated after the conversion of each pixel from thermal inertia to particle size and not simply converting the average thermal inertia as the conversion is non-linear. These thermal inertia and particle size values were calculated using the maximum extent of the Bagnold dune field, while attempting to exclude areas subject to sub-pixel mixing of basement bedrock. This result from the average Bagnold dune field contrasts with our orbitally-derived particle sizes (using the same methods) of Namib dune which range from 312 µm to 438 µm with an average of 396 ± 49 µm (Figure 2). The increased particle sizes at Namib dune and the maximum values of the main Bagnold dunes field likely represent the effect of subpixel mixing with the basement Murray and Stimson formations which commonly have thermal inertia values > ~350 J m$^{-2}$ K$^{-1}$ s$^{-1/2}$ translating to effective particle sizes of >1 mm. This type of mixing is observed at scales (e.g. Figure 1B) finer than the 100m/px THEMIS images and has a dramatic effect on the average derived particle sizes and thermal inertias.

4.2 Particle Sizes Derived from Landed Observations

*4.2.1 GTS Derived Particle Sizes*

As described above, several different scenarios were considered to model the GTS data and derive particle sizes. The first being a simple model of a uniform thermal inertia using the geometry of a best fit plane, and a second, complex facet-based model incorporating small scale topography. The RMS best fit for both the plane and facet models resulted in a thermal inertia of 200 J m$^{-2}$ K$^{-1}$ s$^{-1/2}$. The second fitting metric (minimum average and ΔT temperature differences) for the facet model resulted in a thermal inertia of 170 J m$^{-2}$ K$^{-1}$ s$^{-1/2}$, while the plane model resulted in a thermal inertia of 180 J m$^{-2}$ K$^{-1}$ s$^{-1/2}$. By eye, we estimated the thermal inertia to be 190 J m$^{-2}$ K$^{-1}$ s$^{-1/2}$. This range of thermal inertias translate to particle sizes of ~106-216 µm and are functionally equivalent to particle sizes derived from orbit (within the measurement and model uncertainty). Neither the plane or facet models matched the time of peak temperature recorded by the GTS (~30-45-minute mismatch); however, the facet model typically did a better job at matching the time of peak temperature, with only a ~15-minute discrepancy.

The albedo used in these models was not derived from another dataset, as is common practice from orbit which reduces the number of free parameters, but was instead permitted to vary. This additional free parameter resulted in a significant range of best fit albedos (0.07-0.13), all of which are below the orbitally derived value of ~0.15±0.01 for the main Bagnold dune field.

### *4.2.2 Particle sizes from Imagery*

Particle sizes determined from MAHLI imaging are treated as ground-truth, especially those determined manually. There are significant target-to-target variations in the derived particle sizes determined manually, for example the target Barby is significantly coarser (~320-350 µm) than Otavi or the Gobabeb scoop sites (~150-200 µm). This is likely due to the sampling location with Barby, which was undisturbed near the ripple crest of the stoss side of High Dune (nearby to Namib, Figure 1B), while the other samples, including the Goabeb Scoop Site 2 discard (sieved) and Otavi were acquired from the lee side of Namib dune proper. The grain size differences between the lee and stoss sides of ripples/dunes are expected based on terrestrial experience and as such likely samples a range of grain sizes expected of the active Bagnold dunes. The values determined for the Gobabeb Scoop Site 2 discard and Otavi are most useful for direct comparison to GTS derived particle sizes as the GTS footprint observed the area near to the primary sampling region of Curiosity for the Bagnold dunes campaign (Figure 9).

The particle size values determined automatically via the wavelet method have significantly larger standard deviations than those measured manually, often more than ±100 µm, which limits their utility to discriminate materials with approximately the same particle sizes, in this case fine to medium sands. If these data were used solely without manual point counts to verify the derived particle sizes in the cases measured here, the automated method would likely overestimate the actual particle size (Table 1; Figure 8).

5 *Discussion*

5.1 Cross Dataset Comparisons

To first order, all the datasets examined in this work agree with one another, and for the first time provide a quantitative linkage from measured particle sizes to orbital data of an active dune field on Mars. Fine scale imagery resulted in particle sizes of ~150-350 μm (depending on the sampling location), the GTS data resulted in values ranging from ~110 to 220 μm (considering the range of fitting and modeling methods) and the orbital THEMIS data resulted in particle sizes of ~160 to 340 (from the average and 1-σ values).

However, there are important discrepancies. A much larger range of thermal inertia values was determined from orbit for both the main dune field and Namib dune, likely representing sub-pixel mixing of the higher thermal inertia basement Murray mudstone and Stimson sandstone (Vasavada et al., 2017). This would artificially raise the upper end of derived thermal inertia values which is not adequately captured in the modeling from orbit, and thus bulk thermal inertia interpretations can represent a significant problem for the fine-scale interpretation of future landing sites, as well as geologic contexts. A large range in visible imagery-derived particle sizes was also observed (Figure 8 & Table 1) that likely represents some amount of variability expected to be present over much of the dune field. This variability did not seemingly have a large effect on the derived thermal inertia values, which is discussed further in the next section.

The ~30 J m$^{-2}$ K$^{-1}$ s$^{-1/2}$ differences in the GTS derived thermal inertia values, depending on the type of fit conducted, are likely within or close to the uncertainty of both the instrument and model (i.e., input parameters), and is also commonly expected from orbit.

However, the derivation of both thermal inertia and albedo from a single diurnal curve can also provide non-unique results when considering measurement uncertainties. While the average daily temperature is primarily controlled by the albedo (Figure 6B), thermal inertia plays a secondary role (Figure 6A). The opposite is true for the diurnal temperature range, where thermal inertia is the primary controlling factor, but albedo plays a secondary role. As such the two variables are not completely independent and cannot be perfectly separated given measurement uncertainties when solving for both using a thermal model. The comparison of a thermally derived albedo to an independent albedo determination from additional instrumentation acquired contemporaneously is critical to reducing the free model parameters and providing better model fits to complicated datasets.

5.2 Using Ground-truth to Interpret Orbital Data

GTS provides a critical link between ground-truth particle sizes and the globally available, orbitally derived THEMIS thermal inertia covering ~100% of the Martian surface at 100m/px from 60˚S to 60˚N. As is observed in other regions of Gale Crater (Vasavada et al., 2017), the complex geological surfaces examined by GTS often result in non-unique interpretations. Thermal inertia can be even more difficult to constrain from orbit, with limited context, fewer constraints (typically full diurnal coverage is not available), and larger footprints covering $10^4$ m$^2$ at best with current instrumentation (e.g. THEMIS, TES, etc.). While sand dunes are likely the most thermophysically straightforward features available on the surface of Mars, they still present significant complications. Figures 4A and 5 illustrate the complex surfaces observed in Navcam imagery where ripples are present at cm- to m-scales. Ripples such as these are observed elsewhere on the dune field (e.g. Lapotre et al., 2016) and they complicate the distribution of surface slopes and azimuths. Furthermore, as discussed earlier, the variability in particle size across different sections (e.g., lee, stoss, trough, peak, etc.) of the dune further complicates interpretations. However, the GTS scale footprint permits the assessment of the effects of uncertainties on the thermal data, and thus, due to the scale of the dunes observed here, permits a translation of the GTS findings to orbit for use with interpreting THEMIS and TES data.

We found that the inclusion of small scale ripples (adding a slope and slope azimuth distribution) in the thermal modeling of surfaces did not result in significantly better fits to the measured surface temperatures (Figures 4 & 11). Some important differences are apparent in Figure 11, namely related to the timing and fit of the post-dawn data, which are better modeled when adding the distribution of sub-pixel slopes and slope-azimuth facets. This has important implications for orbitally derived thermal inertias (typically derived pre-dawn) where no differences are observed between simple planar and facet model for fine-scale slopes of the magnitude observed at Namib dune. This result indicates (and is borne out by our measurements) that the thermal inertia of the dune derived from orbit is likely to be representative of the actual dune particle size and does not require the consideration of fine (10s of cm scale) ripple features likely to be prevalent on the surface of dunes.

The second consideration is the potential for armoring and/or induration of the upper ~mm to ~cm of Martian dunes. This also does not likely represent a significant impediment to orbitally derived thermal inertia values, especially for active Martian dunes where relatively thin and only moderately coarser particle size armors have been observed (e.g. Fergason et al., 2006a). While we were able to achieve slightly better fits (e.g. better match the shape of the curve, and not just the nighttime or min/max temperatures) by using a more complicated layered model, the effects, at least in the case of Namib dune, are also very minor. By using the two measured particle sizes from Barby (the armoring coarser layer, ~350μm) and Gobabeb (~150μm) as endmembers in this layered model and varying the thickness of the layer (assuming an albedo of 0.11), we find that layers ~a few mm (up to ~5 mm) thick behave almost identically to the underlying material (Figure 10). A moderately coarse armor (such as that modeled here) that is less than ~1 mm causes a maximum deviation of ~2K at local noon, which is within typical instrument uncertainties reported for THEMIS (Christensen et al., 2004) and GTS (Gómez-Elvira et al., 2012; Sebastian et al., 2010; Vasavada et al., 2017). These coarser or indurated upper layers are more efficient at translating the heat to the insulating base layer, which then controls the overall surface temperature (Figure 10). A surface that has an armoring layer that is thicker than ~30 mm would be dominated by the material properties of the upper layer and behave nearly the same a non-layered surface composed of the same materials. However, armoring layers as thick as this have not been observed on any bedforms by landed spacecraft, therefore these results suggest that the thermal inertia derived from thermal observations likely represent the bulk, underlying material and not the typically thin armoring material. These results are consistent with observations of Saber bedform at the MER Spirit landing site, which was composed of a finer-grained material covered by a coarser-grained armor. The modeling of Mini-TES data found that a thin layer of coarser particles (e.g. 1-2 mm) did not significantly affect the derived thermal inertial of aeolian bedforms. Similarly to the work presented here, the surface temperature of the bedform was controlled primarily by the grain size of the substrate and not the armoring lag (Fergason et al., 2006a). The fits of these significantly more complicated models (both planar and faceted) do not affect pre-dawn temperatures significantly (~1-2K) from a simple planar model, though they do in general fit the rise and late afternoon/evening data better than a simple model (Figure 11). Similarly, this has important considerations for the determination of particle sizes from orbit as the derived thermal inertia, even with some armoring/induration (up to ~5 mm for the particle sizes modeled here), is likely representative of the bulk particle size of the dune and wind regime, rather than the thermal inertia of the lag, which is again borne out by the THEMIS observations shown here.

5.3 Dunes and Sands on Mars

Thermophysical data from THEMIS and other instruments has enabled the determination of the approximate bulk particle size of materials that comprise dunes on Mars (Christensen, 1983; Edgett & Christensen, 1991; Fenton & Mellon, 2006; Fergason et al., 2012; Fergason et al., 2006b). This has improved our understanding of the range of materials mobile on Mars both now and in the recent past. It is instructive to compare the particle size observed at Bagnold with those of other dunes observed globally, both from orbital and surficial measurements, to understand how representative the Bagnold dune field is to other regions on Mars. This comparison can then inform how the results presented here can be applied more broadly to interpret the recent geologic and aeolian history of not only Gale crater, but of the Martian surface.

Edgett and Christensen (1991) determined the average Martian dune particle size of km-scale dune fields (Kaiser, Proctor, Rabe, and Moreux craters) using Viking Infrared Thermal Mapper (IRTM) and Viking Orbiter images. These dunes have thermal inertia values (converted to SI units using a scale factor of 41.86 from Viking-era units of $10^{-3}$ cal cm$^{-2}$ K$^{-1}$ s$^{-1/2}$, and particle sizes updated from thermal inertia values following the same methods from orbit used in this work and assuming a 6 mbar atmosphere) ranging from 326 J K$^{-1}$ m$^{-2}$ s$^{-1/2}$ to 356 J K$^{-1}$ m$^{-2}$ s$^{-1/2}$, with an average thermal inertia of 343 J K$^{-1}$ m$^{-2}$ s$^{-1/2}$. These thermal inertia values correspond to particle sizes greater than ~1000 μm (coarse sand to very coarse sand) with our updated thermal inertia to particle size conversion controlled by lab experiments and numerical models (e.g. Piqueux & Christensen, 2009a, 2009b; Presley & Christensen, 1997b, 1997c) as compared to previous conversion estimates of ~500μm (Edgett & Christensen, 1991). Edgett and Christensen (1994) expanded the work of Edgett and Christensen (1991) by examining IRTM thermal inertia data for dark crater floor units found in latitudes ±55°, correlating thermal inertia values with surface morphology, and interpreting the sedimentary and aeolian transport history of the materials. The average thermal inertia for dark intra-crater features was 270 J K$^{-1}$ m$^{-2}$ s$^{-1/2}$, and thermal inertia values ranged from 160 J K$^{-1}$ m$^{-2}$ s$^{-1/2}$ to 440 J K$^{-1}$ m$^{-2}$ s$^{-1/2}$. Fenton et al. (2003) derived particle size values from TES-derived thermal inertia values for the Proctor crater dune. The average TES-derived thermal inertia value is 277 ± 17 J m$^{-2}$ K$^{-1}$ s$^{-1/2}$, corresponding to an effective particle size of 600 ± 140 μm. Fenton and Mellon (2006) also calculated particle size values from TES-derived thermal inertia values for the Proctor crater dune field. TES thermal inertia values range between 260 J K$^{-1}$ m$^{-2}$ s$^{-1/2}$ and 360 J K$^{-1}$ m$^{-2}$ s$^{-1/2}$, corresponding to particles sizes ranging between ~450 μm and >1 mm (medium to coarse sand). These estimations of dune particle sizes are all comparable when considering the high uncertainties associated with these calculations (Fergason et al., 2006b).

The thermal inertia and particle sizes of bedforms was also determined at both the Mars Exploration Rover (MER) landing sites using Mini-TES data (Fergason et al., 2006a). In many cases, Microscopic Imager (MI) data were also acquired of these features, enabling the direct comparison of particle sizes derived from thermal inertia and imagery in a manner similar to our work at Namib dune. At Meridiani Planum in the bottom of Endurance crater, Mini-TES footprints from the Opportunity rover did not fall cleanly on the stoss or lee side of the bedforms, and therefore all Mini-TES measurements were averaged to determine the thermal inertia of the entire bedform. The bedform at Endurance crater has a thermal inertia of 200 J K$^{-1}$ m$^{-2}$ s$^{-1/2}$, which corresponds to a particle size of ~150 μm (fine sand). The sand at this location was deemed too hazardous to image directly with MI, therefore a different patch of sand closer to the rover was instead imaged. Given the proximity, it was assumed that this sand was likely similar to the bedform material observed by Mini-TES, as such making a direct comparison between MI and Mini-TES reasonable. The particle size of the sand was measured by hand in this location and was ~130 μm in diameter, similar to the values that were thermally derived from Mini-TES (Fergason et al., 2006a) and likely within the measurement/model error.

At the Gusev site, Mini-TES on the Spirit rover observed a bedform in Bonneville crater that consisted of two distinct sections, which were each modeled separately. The upper section was observed to climb the crater wall to the north and had a Pancam (Bell et al., 2008) albedo of ~0.18. The second portion of the observed bedform was located on the lower crater floor and had a Pancam albedo of ~0.23. The thermal inertia derived from Mini-TES for the north crater wall bedform was 200 J $K^{-1}$ $m^{-2}$ $s^{-1/2}$, corresponding to a particle size of ~150 μm (fine sand) (Fergason et al., 2006a). The bedform near the crater floor has a Mini-TES-derived thermal inertia of 160 J $K^{-1}$ $m^{-2}$ $s^{-1/2}$, corresponding to finer particle size of ~60 μm (silt) (Fergason et al., 2006a). Rather than being composed of different sands, Fergason et al. (2006a) suggested that the variable thermal inertia between these two units may instead be caused by a thin layer (≤1 cm) of mantling aeolian dust. This hypothesis is consistent with both the higher albedo and lower thermal inertia values observed. Fergason et al. (2006a) further suggested that the darker particles of the bedform section along the wall were consistent with enhanced bedform activity and as such a less dusty surface. In our above work, we suggested that armoring and/or induration of the upper ~cm of Martian dunes is not a significant factor in the derivation of thermal inertia. The distinction with the observations of the bedforms in Endurance crater is that they were mantled in dust, rather than a coarse-grained armor. The finer-grained dust, given its smaller diurnal skin depth, has a much larger effect on the derived thermal inertia (both from ground and orbital measurements) than a coarser-grained armor, with a larger diurnal skin depth.

The thermal inertia derived for bedforms observed at the MER and MSL sites (i.e., Bonneville and Endurance craters at MER and Bagnold dune field at MSL) are all lower than thermal inertia values derived for intracrater bedforms from orbital IRTM-, and TES-derived thermal inertia and particle size values (e.g. Aben, 2003; Edgett & Blumberg, 1994; Edgett & Christensen, 1991; Fenton et al., 2003). However, the thermal inertia values derived at Bonneville and Endurance craters (~200 J K$^{-1}$ m$^{-2}$ s$^{-1/2}$) are the same as the thermal inertias derived at the Bagnold dunes. The Bagnold dunes are likely to be more active given their low albedos as compared to the Bonneville and Endurance crater bedforms, where optically thick, but thermally thin dust may be present (e.g. Fergason et al., 2006a). Given the modeling findings above and previous results we conclude (e.g. Fergason et al., 2006a)it is likely that the Bonneville and Endurance dunes share a similar bulk particle size as the Bagnold dunes as thin and patchy dust doesn't dramatically affect the thermal inertia of these surfaces, and may have been transported under similar regimes.

The large particle sizes determined in early orbital studies (e.g. coarse to very coarse sands) are likely the direct function of the large measurement spot sizes used (km-scale), rather than related to the bedforms themselves. As shown in this work, the small-scale (cm- to m) features of bedforms do not dramatically affect the effective particle sizes (Figure 4b & c), and rather it is the sub-pixel mixing with the substrate (significantly enhanced as pixel sizes grows) that dominates the uncertainty in particle size. While there is likely some variability among dune particle sizes on Mars, early studies that rely on TES results almost certainly overestimate the particle size for all but the largest dune fields.

*6   Conclusions*

   The Bagnold dune field offers a unique opportunity to directly compare ground-truth particles sizes derived from imagery and lander-scale thermophysical data to orbital data at high enough spatial scales to resolve many Martian dune fields at geologically relevant scales. The cross-comparison of particle size from the significantly different methods presented in this work yield consistent results. This work shows that particle size determination of homogeneous materials, despite the small-scale (i.e., cm- to m-scale) features and thin armoring/induration of the upper surface (i.e., <1 cm), do not dramatically affect the thermal inertia derivation and thus effective particle size determination. Instead, we find that sub-pixel mixing of the substrate (e.g. bedrock) dominates the particle size uncertainty, which can be reduced by acquiring data at finer spatial scales. This effect likely drove the early interpretation of relatively coarse dune particle sizes on Mars (coarse to very coarse sands) due to the reliance and availability of km-scale TES and IRTM data. We propose that dune particle sizes on Mars are likely more similar to those observed in the Bagnold dunes (fine-medium sands), rather than those derive from early orbital studies.

## 7  Acknowledgements

The authors acknowledge the MSL team who operates the Curiosity rover on Mars, as well as the REMS instrument team, that developed the Ground Temperature Sensor. The authors thank Daniel Buscombe for assistance with the pyDGS software package. Data used in this work was obtained from the NASA Planetary Data System. Additional reduced data products are available from Christopher Edwards ([Christopher.Edwards@nau.edu](Christopher.Edwards@nau.edu)), NAU's Open Knowledge website ([http://openknowledge.nau.edu/id/eprint/5230](http://openknowledge.nau.edu/id/eprint/5230)), modeling and data analysis tools are freely available ([http://krc.mars.asu.edu](http://krc.mars.asu.edu) and [http://davinci.asu.edu](http://davinci.asu.edu)). This research was funded by the MSL Participating Scientist program and the 2001 Mars Odyssey THEMIS instrument. Part of this work was performed at the Jet Propulsion Laboratory, California Institute of Technology, under a contract with NASA.

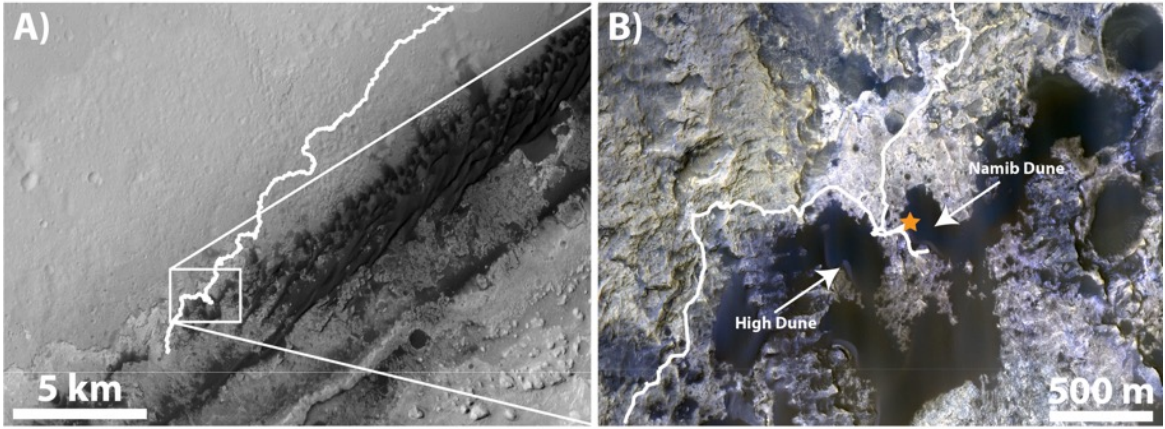

***Figure 1. (A)*** *Context Image mosaic with the rover traverse overlain,* ***(B)*** *HiRISE color mosaic of the Bagnold dunes area. The orange star corresponds to the location where activities associated with Namib dune from Sol 1222-1242 were undertaken.*

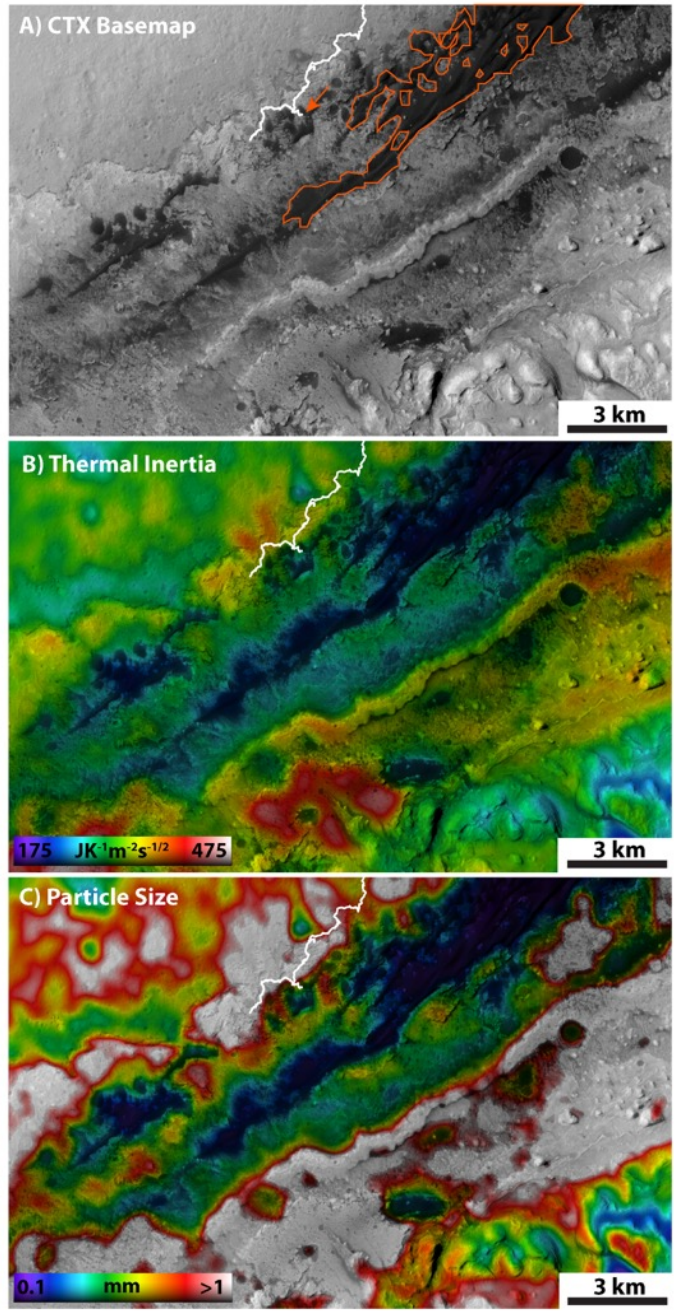

***Figure 2.*** *(**A**) CTX basemap of lower Mount Sharp, including a portion of the rover traverse path, the outline of the Bagnold dunes corresponding to our grain size results, and an arrow that identifies Namib dune. (**B**) THEMIS thermal inertia derived from I17950012 of lower Mt. Sharp colorized and overlain on the CTX basemap. (**C**) THEMIS derived particle size from I17950012 of lower Mt. Sharp colorized and overlain on the CTX basemap. Particle sizes >1mm are not distinguishable and are colorized as white.*

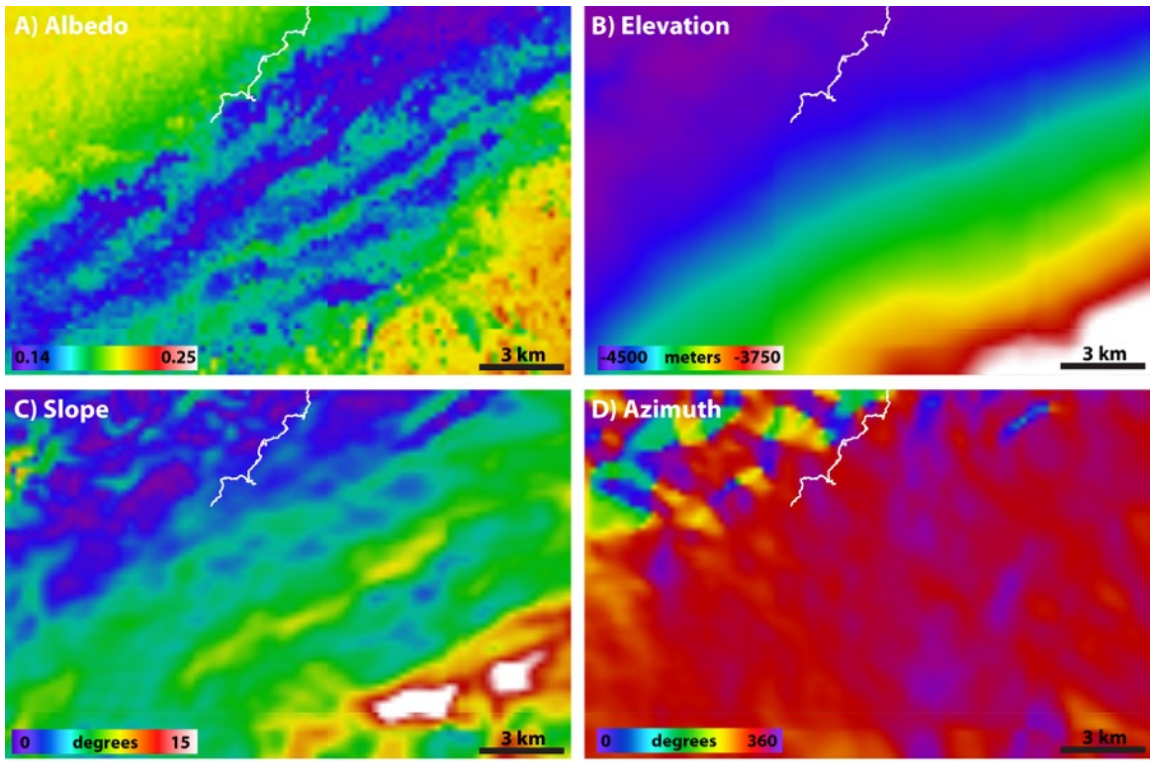

*Figure 3. (A)* THEMIS derived visible Lambert albedo scaled to 100m/px (Edwards et al., 2011a), *(B)* CTX derived elevation scaled to 100m/px, *(C)* CTX derived slopes scaled to 100m/px, and *(D)* CTX derived slope azimuth scaled to 100m/px used in the derivation of the thermal inertia (Figure 2B) and ultimately particle size (Figure 2C).

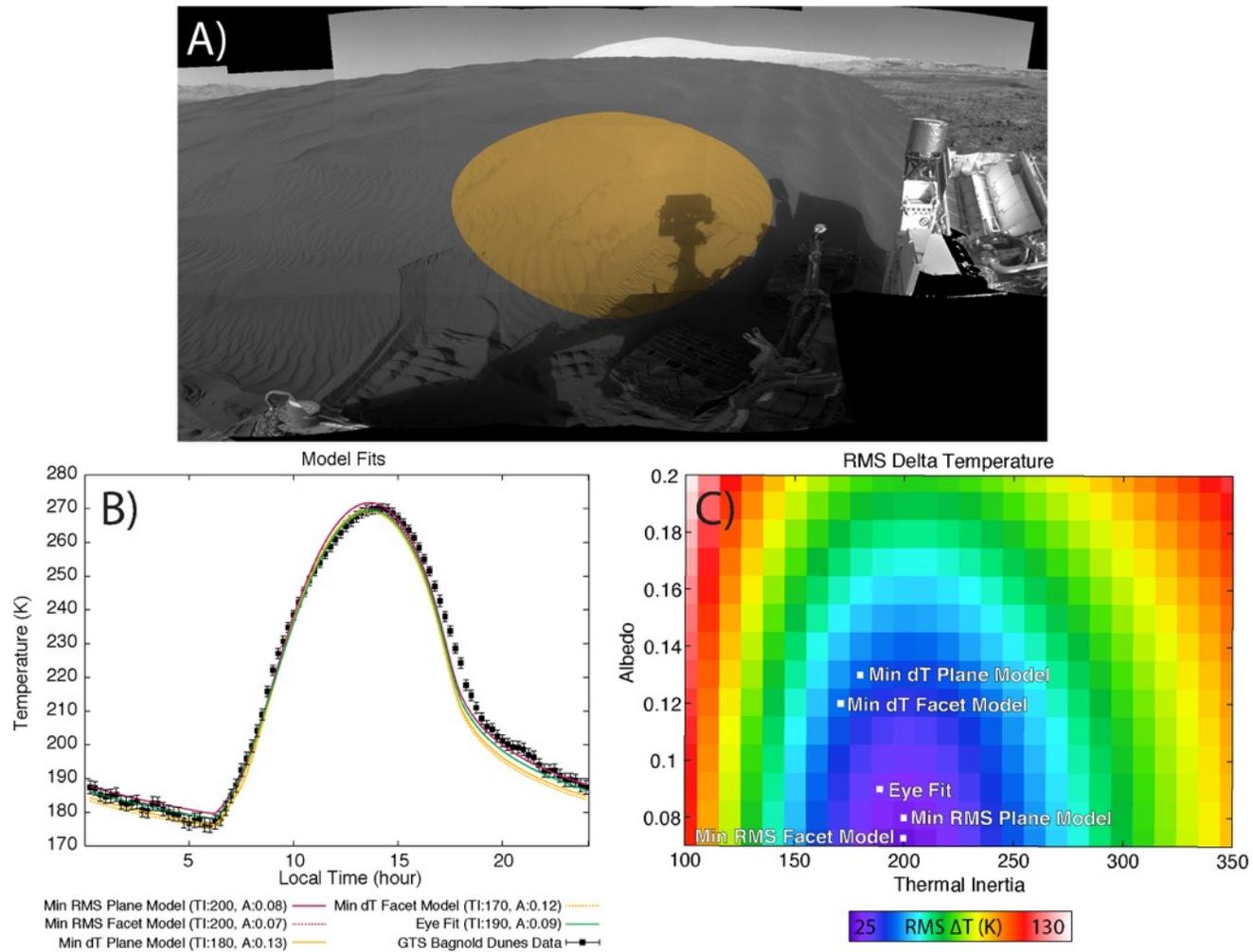

*Figure 4. (A)* Navcam mosaic of Namib Dune. The approximate GTS footprint is highlighted in orange *(B)* GTS data acquired over sols 1222-1242 along with several different model fits. *(C)* Difference of modeled and measured RMS temperatures
 for a range of albedo and thermal inertias. The locations of the model fits in *(B)* are shown.

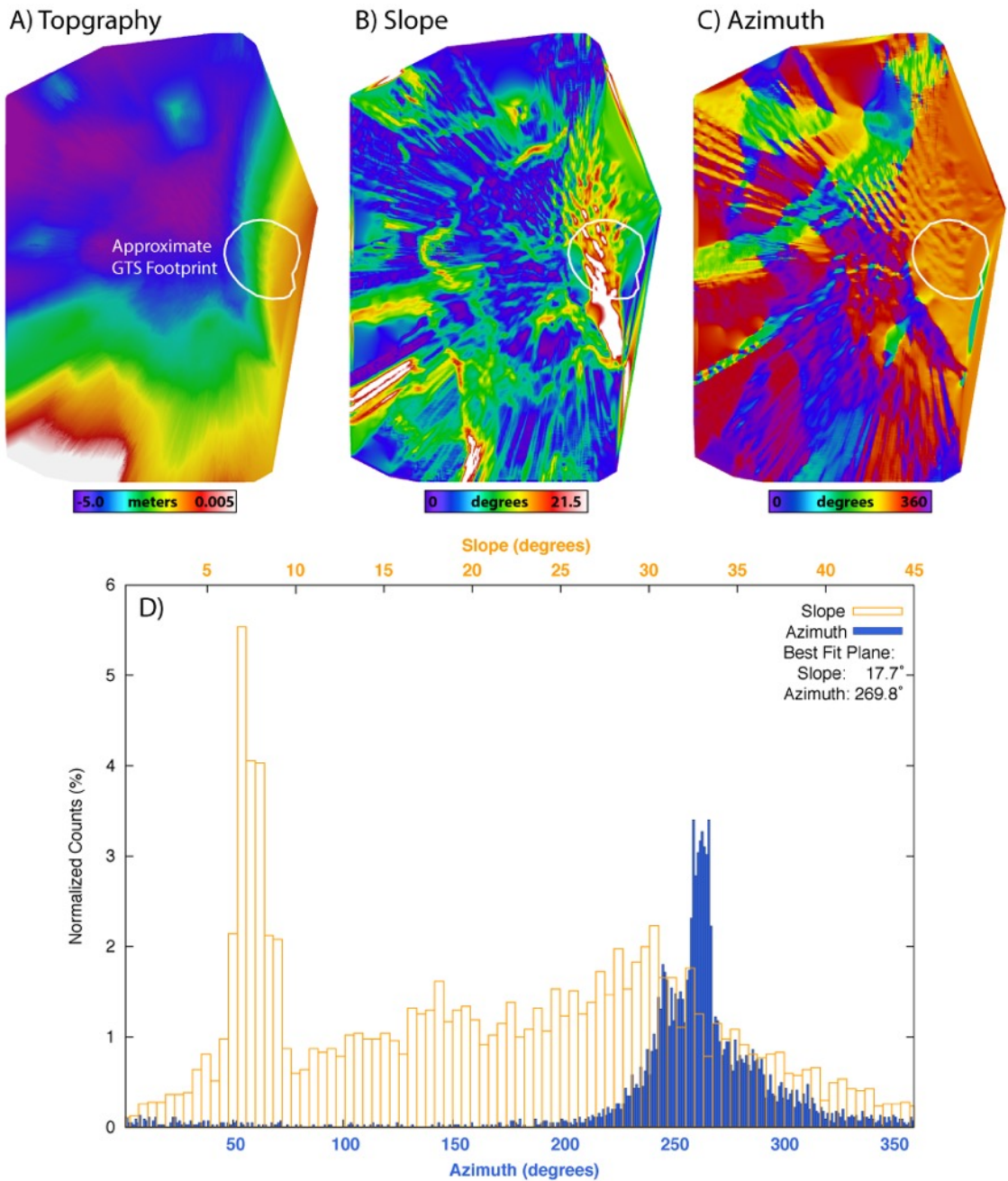

*Figure 5. Navcam derived **(A)** topography, **(B)** slopes, **(C)** slope azimuth of Namib Dune. The approximate GTS footprint is highlighted in by the white outline. **(D)** A normalized histogram showing the distribution of slopes and azimuths under the GTS footprint. A best fit plane to the GTS footprint results in a slope of 17.7° and azimuth of 269.8°.*

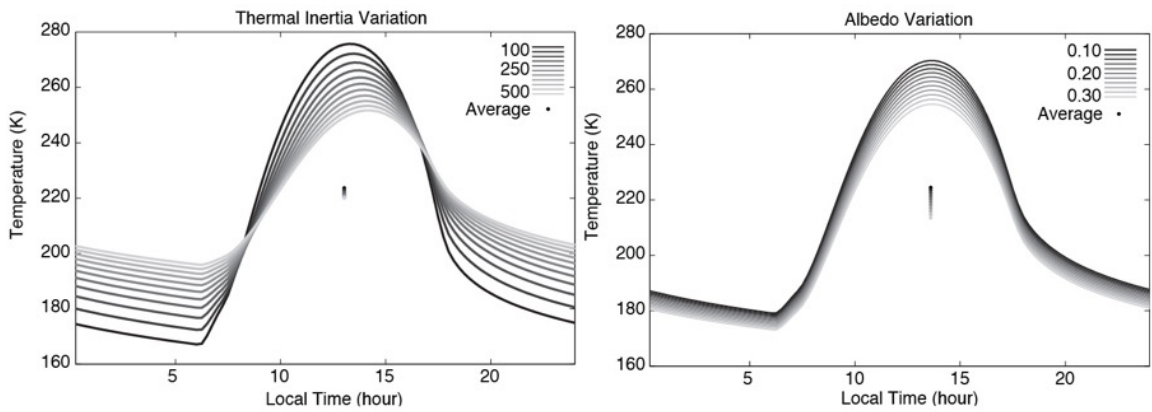

***Figure 6.*** *Effects of **(A)** thermal inertia and **(B)** albedo over the ranges used in the model fits. While the average diurnal temperature (dots) primarily depends on the albedo of the surface, thermal inertia contributes as a second order effect complicating the method of using average temperature to derive a "thermal" albedo.*

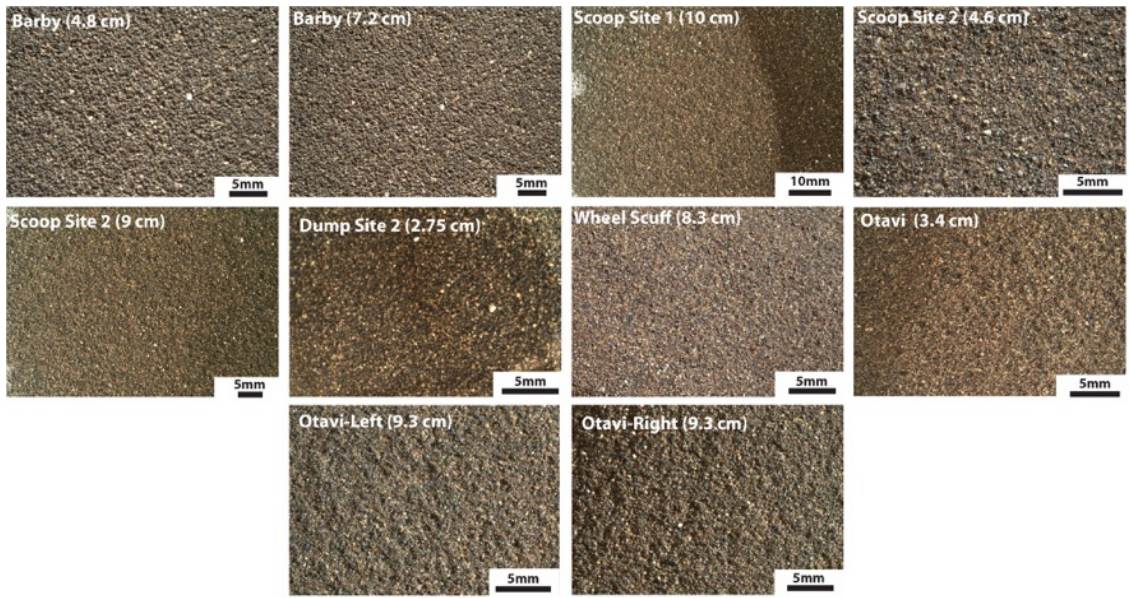

*Figure 7.* MAHLI images at a variety of standoffs taken over the campaign at Namib dune.

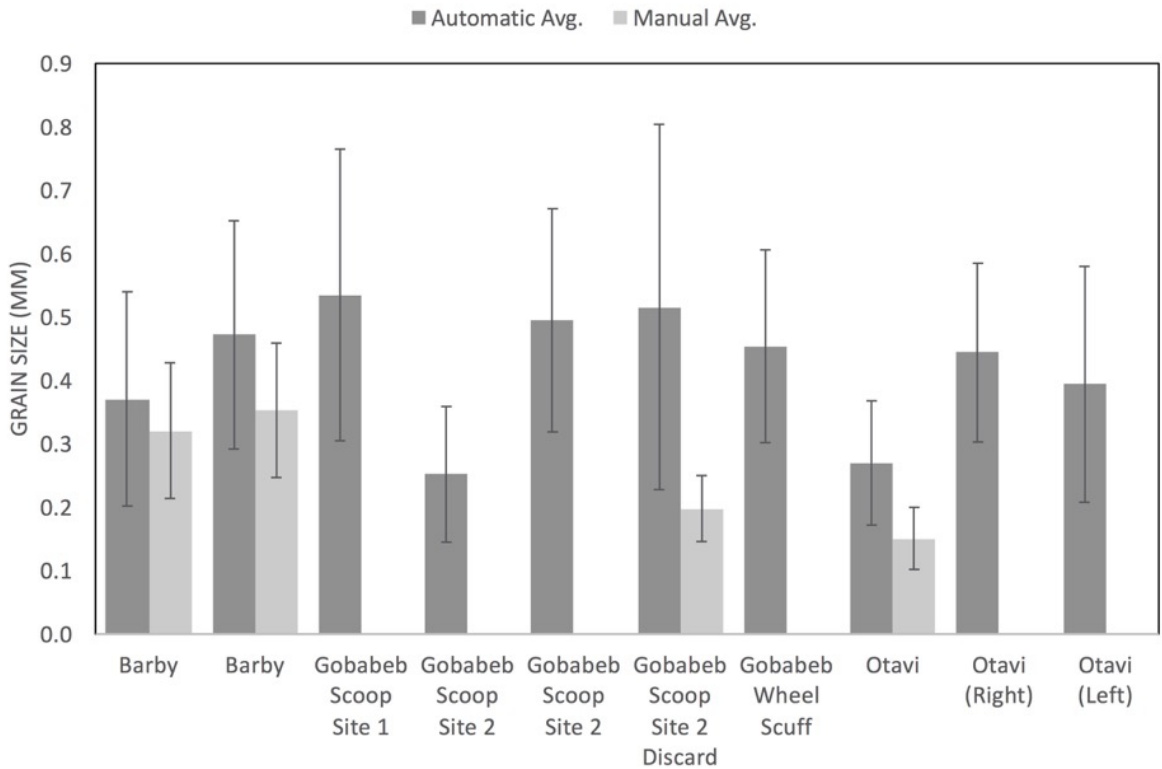

*Figure 8.* Bar graph of the average particle sizes determined from both by the Digital Particle size analysis methods of Buscombe (2013)(dark gray) and by manual point counts (light gray). Averages are shown with 1 sigma standard deviations. Manual point counts have ~100-300 measurement location each.

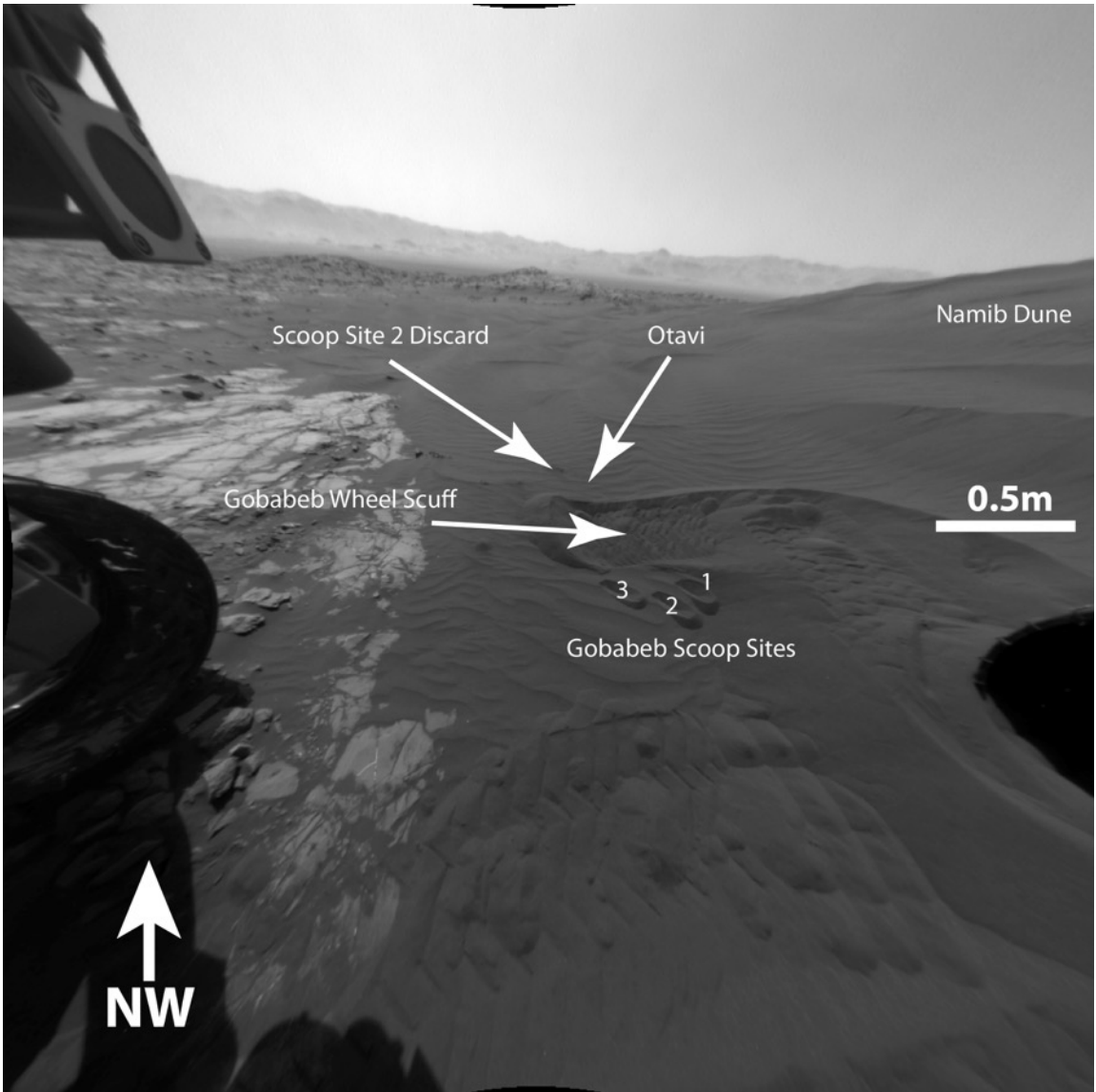

*Figure 9.* Sampling location at Namib dune shown in the front left hazard camera. Highlighted locations correspond to areas where MAHLI images in Table 1 and Figure 7 were acquired.

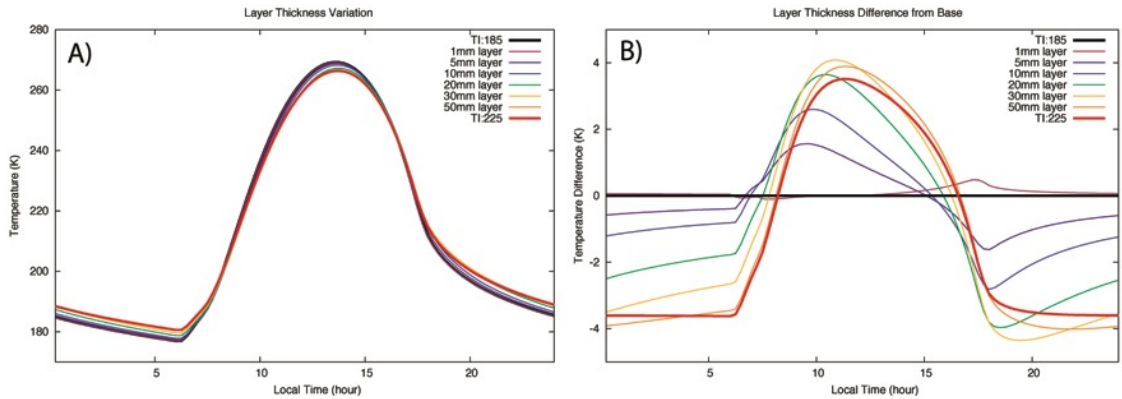

*Figure 10.* Thin layers of coarse or cemented material (225 J K$^{-1}$ m$^{-2}$ s$^{-1/2}$, derived from MAHLI particle sizes of Gobabeb) are modeled on top of a thermally thick layer of lower thermal inertia (185 J m$^{-2}$ K$^{-1}$ s$^{-1/2}$, derived from MAHLI particle sizes of Barby). Layers of coarser material up to ~a few mm behave identically to the thermally thick subsurface materials. Layers up to ~20mm are likely to be indistinguishable in GTS and orbital data as the difference from a lower thermal inertia surface would be less than a few K. Layers greater than a few cm (~30 mm) thick are dominated by the material properties of the upper layer and behave nearly the same a non-layered surface composed of the same materials. Surfaces with these characteristics would be indistinguishable within instrument temperature uncertainties typical for GTS and THEMIS. Full diurnal curves are shown in *(A)*, while the difference from the 185 J K$^{-1}$ m$^{-2}$ s$^{-1/2}$ material are shown in *(B)*.

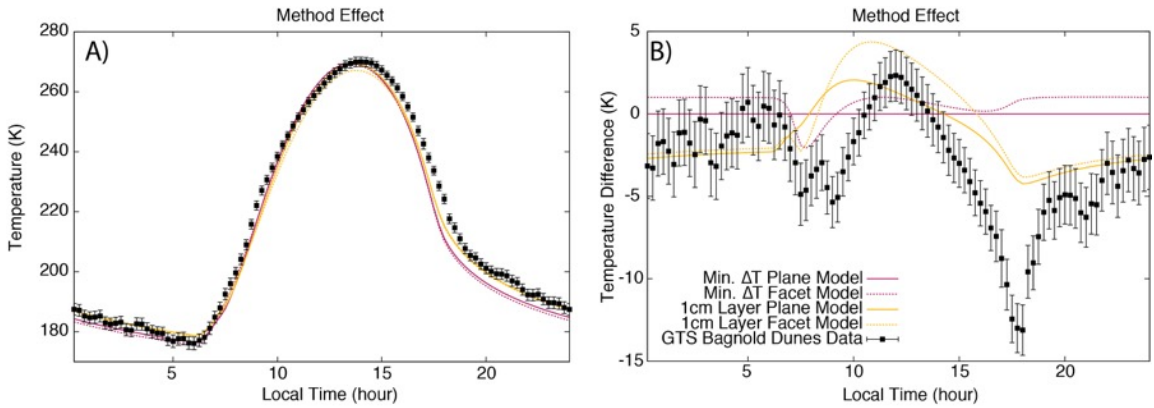

*Figure 11.* *The best fit ΔT plane (180 J m$^{-2}$ K$^{-1}$ s$^{-1/2}$; 0.13 albedo) and facet model (170 J m$^{-2}$ K$^{-1}$ s$^{-1/2}$; 0.12 albedo) data are compared to a layered plane and facet model with a 1 cm layer of 350 μm particles with an albedo of 0.12 on top of a 150 μm particle size substrate. Full diurnal curves are shown in **(A)**, while the difference between a simple planar model and the other scenarios are shown in **(B)**. In general, the most complicated model (faceted and layered) provides the best fit of the data, matching the early morning data, post-sunrise trends and the evening (~7 pm) data. No model used in this work is able to accurately model either the peak temperature timing or the mid-afternoon temperature drop, indicating there are some additional physical properties not being adequately addressed.*

**Table 1.**

| Target | Derived Standoff | Resolution (mm/px) | Grain Sizes (mm) | | | | Image ID |
|---|---|---|---|---|---|---|---|
| | | | Auto. Avg. | Stddev | Manual Avg. | Stddev | |
| Barby | 4.80 | 0.024 | 0.371 | 0.169 | 0.321 | 0.107 | 1184MH0001630000402969R00 |
| Barby | 7.20 | 0.032 | 0.472 | 0.180 | 0.353 | 0.106 | 1184MH0001630000402971R00 |
| Gobabeb Scoop Site 1 | 10.00 | 0.042 | 0.535 | 0.230 | | | 1224MH0001700000403204R00 |
| Gobabeb Scoop Site 2 | 4.60 | 0.023 | 0.252 | 0.107 | | | 1228MH0001630000403298R00 |
| Gobabeb Scoop Site 2 | 9.00 | 0.038 | 0.495 | 0.176 | | | 1224MH0001700000403200R00 |
| Gobabeb Scoop Site 2 Discard | 2.75 | 0.017 | 0.516 | 0.288 | 0.198 | 0.052 | 1242MH0005620020403674C00 |
| Gobabeb Wheel Scuff | 8.30 | 0.036 | 0.454 | 0.152 | | | 1228MH0001630000403296R00 |
| Otavi | 3.40 | 0.019 | 0.270 | 0.098 | 0.151 | 0.049 | 1242MH0005740000403707R00 |
| Otavi (Right) | 9.30 | 0.039 | 0.444 | 0.141 | | | 1242MH0005740000403719R00 |
| Otavi (Left) | 9.30 | 0.039 | 0.394 | 0.186 | | | 1242MH0005740000403719R00 |